\documentclass[a4paper]{article}
\usepackage[linesnumbered,boxed,ruled,commentsnumbered]{algorithm2e}
\usepackage{config/INTERSPEECH2022}
\usepackage{amsthm,amsmath,amssymb}
\usepackage{multirow}
\usepackage{enumerate}
\usepackage{makecell}
\usepackage{cite}
\usepackage{mathrsfs}
\usepackage{float}  
\usepackage{subfigure}  
\usepackage{epsfig}
\usepackage{verbatim}
\usepackage{bigstrut}

\title{The HCCL system for VoxCeleb Speaker Recognition Challenge 2022}
\name{Zhenduo Zhao$^{1,2}${\rm\textsuperscript{*}}\thanks{\textsuperscript{*}equal contribution}, Zhuo Li$^{1,2}${\rm\textsuperscript{*}}, Wenchao Wang${^1}$, Pengyuan Zhang${^{1,2}}$}
\address{
  $^{1}$Key Laboratory of Speech Acoustics and Content Understanding, Institute of Acoustics, Chinese \\ Academy of Sciences, Beijing, China \\
  $^{2}$University of Chinese Academy of Sciences, Beijing, China}
\email{li\_zhuo@foxmail.com,\{zhaozhenduo,wangwenchao,zhangpengyuan\}@hccl.ioa.ac.cn}

\begin{document}

\maketitle
\begin{abstract}
  This report describes our submission to track1 and track3 for VoxCeleb Speaker Recognition Challenge 2022(VoxSRC2022). Our best system achieves minDCF 0.1397 and EER 2.414 in track1, minDCF 0.388 and EER 7.030 in track3.
\end{abstract}
\noindent\textbf{Index Terms}: speaker recognition, VoxSRC2022,  domain adaptation, clustering algorithm, label correction.

\section{System Description for Track1}

\subsection{Data}

\textbf{Training Data:}
  We use VoxCeleb2-dev(vox2dev) \cite{voxceleb2} as training data, which consists of 1092009 utterances from 5994 speakers. To augment data, we first use the SoX speed function with speeds 0.9 and 1.1 to generate extra twice speakers \cite{speed}. In total, there are 17982 speakers and 3276027 utterances. Then, we use MUSAN \cite{musan} and RIRs noises \cite{rirs} to perform online data augmentation. Similar to SpeakIn systems for VoxSRC2021 \cite{zhao2021speakin}, we used a chain augment pipeline to generate samples:
  \begin{itemize}
  \item MUSAN noise with probability 0.2
  \item MUSAN music with probability 0.2
  \item MUSAN speech with probability 0.2
  \item RIRs noises with probability 0.6
  \end{itemize}
  SpeechBrain \cite{speechbrain} was used to build the pipeline. 

\textbf{Developing Data:}
  We use official validation sets \cite{voxceleb1,voxceleb2,voxsrc2020,voxsrc2021} to evaluate our models: Vox1-O, Vox1-E, Vox1-H, Vox20-dev, Vox21-dev and Vox22-dev.

\textbf{Features:}
  We extract 80-dimensional log-Mel Filter Banks (Fbank) as input features without any voice activity detection(VAD). The frame length is 25ms and the frame shift is 10ms. Cepstral mean normalization(CMN) is applied. Our implementation was based on TorchAudio \cite{torchaudio}.

  \begin{table*}[htb]
    \caption{Res2Net50's Performance on VoxCeleb Official Evaluation Sets}
    \label{tab:res2net50result}
    \centering
    \footnotesize
    \setlength{\tabcolsep}{1.2mm}{
    \begin{tabular*}{\linewidth}{lccccccccccccccc}
      \toprule
      \multirow{2}*{Stage} & \multicolumn{2}{c}{Vox1O} & \multicolumn{2}{c}{Vox1E} & \multicolumn{2}{c}{Vox1H} & \multicolumn{2}{c}{Vox20-dev} & \multicolumn{2}{c}{Vox21-dev} & \multicolumn{2}{c}{Vox22-dev} & \multicolumn{2}{c}{Vox22-eval} \\
      \cmidrule{2-15}
      & EER & DCF$_{0.05}$ & EER & DCF$_{0.05}$ & EER & DCF$_{0.05}$ & EER & DCF$_{0.05}$ & EER & DCF$_{0.05}$ & EER & DCF$_{0.05}$ & EER & DCF$_{0.05}$ \\
      \midrule
      base & 0.670 & 0.0483 & & & & & & & & & & & \\
      +LMF & 0.516 & 0.0303 & 0.656 & 0.0386 & 1.172 & 0.0670 & 1.993 & 0.1024 & 1.942 & 0.1091 & 1.751 & 0.0977 & 3.126 & 0.1715 \\
      ++AS-norm & 0.532 & 0.0298 & 0.626 & 0.0370 & 1.123 & 0.0644 & 1.922 & 0.0980 & 1.891 & 0.1047 & 1.726 & 0.0982 & 3.024 & 0.1662 \\
      +++QMF & 0.473 & 0.0283 & 0.587 & 0.0350 & 1.059 & 0.0610 & 1.792 & 0.0972 & 1.795 & 0.1067 & 1.657 & 0.0972 & 2.983 & 0.158 \\
      \bottomrule
    \end{tabular*}}
  \end{table*}
  
\begin{table*}[htb]
  \caption{Experiment results on Vox22-dev and Vox22-eval}
  \label{tab:word_styles}
  \centering
  \begin{tabular*}{\linewidth}{clc p{2cm} p{2cm} p{2cm} p{2cm}}
    \toprule
    \multirow{2}*{Index} & \multirow{2}*{Backbone} & \multirow{2}*{Pooling} & \multicolumn{2}{c}{VoxSRC2022-dev} & \multicolumn{2}{c}{VoxSRC2022-eval} \\
    \cmidrule{4-7}
    & & & EER & minDCF$_{0.05}$ & EER & minDCF$_{0.05}$ \\
    \midrule
    1 & Res2Net50-32 & ASP & 1.657 & 0.0972 & 2.983 & 0.158 \\
    2 & ECAPA-TDNN-large & ASP & 2.402 & 0.1585 & - & - \\
    3 & ECAPA-TDNN-X3 & ASP & 1.930 & 0.1245 & - & - \\
    4 & ECAPA-TDNN-X4 & ASP & 1.912 & 0.1179 & - & - \\
    5 & ResNet34-64 & ASP & 2.112 & 0.1297 & - & - \\
    6 & ResNetSE101-32 & CWC & 1.782 & 0.1091 & - & - \\
    7 & ResNetSE101-64 & CWC & 2.145 & 0.1437 & - & - \\
    8 & Res2Net50-64 & ASP & 2.021 & 0.1356 & - & - \\
    9 & HS-ResNet-DSSA & ASP & 2.082 & 0.1279 & - & - \\
    10 & RepVGG-B1 & ASP & 1.725 & 0.1007 & - & - \\
    \midrule
    fusion \\
    \midrule
    fusion1 & \multicolumn{2}{c}{1+3+4+6+8+9+10} & 1.484 & 0.0873 & 2.585 & 0.1408 \\
    fusion2 & \multicolumn{2}{c}{1+2+3+4+5+6+7+8+9+10} & 1.495 & 0.0874 & 2.538 & 0.1483 \\
    fusion3 & \multicolumn{2}{c}{1+2+3+4+5+6+7+8+9+10+LR} & 1.382 & 0.0825 & 2.414 & 0.1397 \\
    \bottomrule
  \end{tabular*}
\end{table*}

\subsection{Model Structures}
  Two main stream of current most popular model structures was used for the challenge: 1D-convolution-based ECAPA-TDNN \cite{ecapa} and its variants, 2D-convolution-based ResNet \cite{resnet} series and RepVGG \cite{repvgg}. 

\textbf{ECAPA-TDNN}:
  We trained ECAPA-TDNN large with 1024 channels and its two variants, ECAPA-TDNN-X3(EX3) and ECAPA-TDNN-X4(EX4). We used SpeechBrain implementation with 1024 channels. To further boost its performance, we made it deeper and added branches in res2block to enhance representational ability. For EX3, we combine two se-res2block as one basic block. For EX4, we added one more basic block, and meanwhile, to restrict the receptive field, we did not use dilation in the first few blocks. Besides, we used group convolution to preserve the original feature map.

\textbf{ResNet/SE-ResNet}:
  ResNet is one of the most popular model structures currently. Here we used standard ResNet34 with 64 channels. Squeeze excitation module \cite{se} uses an attention mechanism to re-weight feature map channels. Besides, we use a modified version of ResNet as described in \cite{sogou} except that the SE module was added only in the first two blocks. We trained SE-ResNet101 with 32 channels and 64 channels.
  
\textbf{HS-ResNet \& Res2Net}
  In order to model multi-scale features, we used Res2Net \cite{gao2019res2net} and its variants hs-resnet\cite{hsnet} with dssa module. Both models are 50 layers deep, scale 8 and width 14 in Res2Net and scale 8 and width 6 in HS-ResNet.

\textbf{RepVGG}
  RepVGG use 3 branches of convolution and batch normalization when training and re-parameterize them as one for inference. Branch greatly boost capability to model multi-level features. We use the official version with configuration b1.

\subsection{Pooling}
  Deep neural network-based systems use a pooling layer to aggregate frame-level features into segment-level embeddings. We used two pooling methods: attentive statistics pooling(ASP) \cite{asp} and channel-wise correlation pooling(CWC) \cite{cwc}.

\subsection{Loss Function}
  Margin-based loss functions have greatly improved system performance. In this challenge, we adopted circle loss \cite{circle}. Moreover, subcenter \cite{subcenter} and intertopk \cite{intertopk} are two plugin methods that could further improve the discrimination of embeddings.
  Circle loss is formulated as follows:
  $$
    L_{circle} = - log \frac{e^{s \cdot (m^2 - (1-s_p)^2)}}{e^{s \cdot (m^2 - (1-s_p)^2)} + \sum^C_{j=1,j \neq i} e^{s \cdot ((s_n^j)^2)-m^2}}
  $$
  where $m$ is the margin and $s$ is the scale factor.

\subsection{Training Protocol}
  We trained models with a two-stage protocol. All experiments were based on the PyTorch \cite{pytorch}.

  Adam optimizer with weight decay 5e-5 was used in the first stage. Cycle learning rate scheduler \cite{cycle} was adopted, where the minimum learning rate is 1e-8 and the maximum is 1e-3, we trained 2 cycles with one cycle of 100k steps. The batch size was 1024 and the segment duration was 2s. Margin and scale were set to 0.35 and 60 for circle loss, respectively. We used subcenter k=3 and intertopk k=5, m=0.1.

  The second stage was large margin fine-tuning(LMF), we expanded segment duration to 6s, only removed intertopk from loss function, and increased weight decay from 5e-5 to 4e-4. We used a constant learning rate of 2e-5 for 10k steps.

\subsection{Back-end}
  After LMF, cosine distance was used for $4s \times 10$ scoring. Evenly cut 10 4-second long segments from utterances, the mean of score matrix with the size of $\mathbb{R}^{10 \times 10}$ served as the score of a trial. Then, we used adaptive score normalization(AS-norm) \cite{asnorm} and quality measure functions(QMF) \cite{thienpondt2021idlab} to calibrate the scores. For AS-norm, speaker-wise averaged embeddings from vox2dev, leading to 5994 cohort speakers with top 400 imposter scores were used. By the way, we removed imposter variance. For QMF, we followed IDLAB's method to generate 30k trials from vox2dev. Then we trained the logistic regression(LR) model to calibrate the AS-normed score. In the end, another LR model was used to combine all of the calibrated models to get the final fused score. While the generated trials could not perfectly fit the evaluation distribution, we manually tuned the model weights based on Vox22-dev trials.

\subsection{Results}
\subsubsection{Ablation Study}
  Res2Net50 is our best single system, table 1 shows its performance on all of the evaluation sets at different stages. Equal Error Rate(EER) and minimum Decision Cost Function(minDCF) with $C_{FA}=1,C_M=1,P_{target}=0.05$ was reported. After the first stage training, model achieves $EER=0.67\%,minDCF=0.0483$ on Vox1O. EER improved from 0.67\% to 0.516\% and minDCF improved from 0.0483 to 0.0303 after LMF. AS-norm further decreased EER to 0.532 and minDCF to 0.0298. QMF finally push the limit of the model to EER=0.473\% and minDCF=0.0283 . In total, with these stacked methods, the performance got relative improvement of 29.4\% and 41.4\% on EER and minDCF, respectively. 

\subsubsection{System Performance}
Table 2 shows our 10 subsystems performance on Vox22-dev and Vox22-eval. We found that for TDNN-based model, the larger model is, the better performance we got. It was half true for ResNet-series models, as we found simply double the channels cannot bring improvement. 

We fused ResNet-series models firstly with equal weights, got EER=2.585, minDCF=0.1408. Adding TDNN-series models only got improvement on EER but degradation on minDCF. Then, we introduced LR model to train on generated QMF set and tuned the weights based on model coefficients, finally achieves EER=2.414 and minDCF=0.1397.

\section{Semi-Supervised Domain Adaptation}
Semi-supervised speaker recognition attempts to automatically exploit a large amount of target or source unlabeled data in addition to a large amount of source or target labeled data to improve performance.
There are three general goals, one is to obtain better performance on the target domain data, the other is to improve the performance on all domains, that is, to improve the domain robustness, and the third is to achieve better performance on the source domain data.
The goal of this competition is to achieve the performance of the target domain.

We attempt two frameworks, one is pseudo labeling, and the other is self-supervised learning.
The pseudo-label solution contains five stages:
\textbf{1} source label data model training,
\textbf{2} embeddings domain adaptation,
\textbf{3} pseudo-label generation,
\textbf{4} Supervised training on target domain data with pseudo-labels and source domain label data,
\textbf{5} pseudo label correction and re-train.
\subsection{Pre-processing}
Because of the lack of filtering when constructing the CN-Celeb2\cite{li2022cn,cnceleb}, it contains much noisy audio. One of the most intuitive manifestations is that there is much-repeated audio in CN-Celeb2, and some are given different labels.
We directly used md5sum to de-duplicate the speech and the number of audio decreased from 455,946 to 409,628.

\subsection{Base Model Training}
To obtain high confidence edges by using the voting strategy, it is beneficial to select models with as much variance as possible, from the model structure to the training Protocol. We selected the following five models:
(1) \emph{SE-ResNet34} with 32 channel; (2)\emph{ECAPA-TDNN} with 1024 channel; (3) \emph{Conformer-MFA} with 256 hidden dim \cite{zhang2022mfa}; (4) \emph{SE-ResNet101} with 32 channel; (5) \emph{Cot-Net}\cite{cot} with 32 channel;
Others settings are shown in Table~\ref{t-base}.

\subsection{domain adaptation}
During the evaluation process, domain adaptation is necessary due to a large domain mismatch.
An important manifestation of domain mismatch on embedding is the difference between the mean and variance. Thus, the simplest approach is to align the centers of the different domains directly, and experiments show significant improvements.
Further, aligning the variance can also achieve improvements in theory. We attempt to apply CORAL \cite{sun2017correlation}, CORAL+\cite{lee2019coral+} and CORAL++\cite{li2022coral++} into embeddings of target domain directly and use cosine similarity for scoring. However, there is no performance improvement unless we use back-ends, LDA, and PLDA\cite{plda,plda02}. 
Due to time constraints and inconvenient operations, we do not do this work systematically and will do these in the future.

\subsection{cluster}
Because AHC is computationally infeasible and k-means depend on the estimate of k, we propose a novel cluster algorithm, a progressive sub-graph clustering algorithm based on two Gaussian fitting and multi-model voting, denoted as GMVPG clustering.
The key points of this algorithm are as follows:
First, finding high-confidence positive trials, that is, edges, using a multi-model voting strategy based on the KNN affinity graph.
Secondly, utilizing connected sub-graphs to obtain pseudo labels
Then, using iterative top-k information to gradually merge sub-classes to prevent super-classes.
Finally, two Gaussian distributions are introduced to fit the intra-class score distribution to further check for high-confidence edges.
The detailed algorithm is shown as follows:

\subsection{Supervised training and fine-tune}
\subsubsection{Training stage1}
The training data for track3 contains VoxCeleb2, the unlabeled target domain dataset from Cnceleb2, and the small amount of labeled set.
Speed perturbation augmentation is used in all data, and other augmentations are the same as track 1.
In the first training stage for track 3,
we explore two training strategies, one is to train the model from scratch, and the other is to utilize models from Track1 as the pre-trained model.
For the latter, it is necessary to make the new model be converged before starting training, freezing the extractor and training the classification layer first is an effective approach.
In addition, we also explore two different training protocols, one is Adam optimizer with cycle learning rate scheduler, as shown in Section 1, and the other is SGD optimizer with ReduceLROnPlateau scheduler.
\begin{algorithm}
    \SetAlgoNoLine 
    \SetKwInOut{Input}{\textbf{Input}}\SetKwInOut{Output}{\textbf{Output}} 
    \Input{
        Embeddings of target domain audio data from multi models after domain adaptation $X^t, t={0,1,...,T}$\;\\
        }
    \Output{
        pseudo label of target domain audio data $D$, here, $Ans = (d_i,y_i),i = {0,1,...,M}, y_i={-1,0,1,...,N}$\;\\
        }
    \BlankLine
    
    Collect information about KNN affinity graph for D, K is set to 500; $sim(x^t_i,k)$ represents the top $k st\ $ similarity\  of\ $x^t_i$\;
    filter out partial audio $d_i$, if $\ \exists t, sim(x^t_i, K) > th_{high}$\;
    Construct initial KNN affinity graph for D, initial k is set to 10\;
    preserve edges between $d_i$ and $d_j$, $e_{ij}$ if $sims(x^t_i, x^t_j) >= sim(x^t_i,k)\quad \forall t$, other edges are deleted\; Utts are deleted if no edges are connected. For convenience, we denote $E_k = {e_{ij}}$ for all preserved edges and $utt_k = {d_i}$ for all preserved utts when $k$, respectively\;
    Obtain initial labels by searching connected sub-graph (SCSG), $G(k)$, one sub-graph means one class\;
    
    \Repeat
        {\text{k=50}}
        {
            $E_{add} = E_{k+5} - E_{k}$, $U_{add} = U_{k+5} - U_{k}$\;
            Split $E_{add}$ to $E_{old.old}$, $E_{new.old}$ and $E_{new.new}$ according to whether utt is in $U_k$\;
            Generate temp pseudo labels $tmpG_{new}(k+5)$ for $U_{add}$ based on $E_{new.new}$ only\;
            Combine class for $U_k$ based on $E_{old.old}$ only; $tmpG_{old}(k+5) = {subspkCombine} (E_{old.old}, G(k))$\;
            Process the relationship between $U_{add}$ and $U_k$; $G(k+5) = {CbNewOld} (tmpG_{new}(k+5), G(k), E_{new.old})$\;
            k = k + 5\;
        }
    Throw out classes with fewer than 10 utts\;
    
\caption{GMVPG clustering algorithm \label{al3}}
\end{algorithm}

\subsubsection{Training stage2}
In this stage, we only use CN-Celeb dataset without speed perturbation to finetune all systems. But, the VoxCeleb weights of the classification layer are preserved to prevent overfitting.
Two fine-tuning training protocols are utilized, one is SGD with a 2e-5 learning rate, and the other is Adam with a 2e-5 learning rate. Others are the same as Track 1.

\subsection{pseudo label correction}
Since the GMVPG method brings some mislabeled and noisy samples, it is important to correct labels after the initial model training.
The key points of this algorithm are as follows:

\textbf{i}. Split audio into three types according to its similarity to centers, high/median/low-confidence;
\textbf{ii}. Stas the correlation between each class based on audio with median-confidence
\textbf{iii}. Integrate the results from multi models to correct the labels.

The detailed algorithm is shown as follows:
\textbf{1}. calculate the similarity of all audio in CN-Celeb to the two most similar class centers, denoted as $sim_{i}^{top1}$ and $sim_{i}^{top2}$, the labels of most similar class centers are denoted as $y_{i}^{top1}$ and $y_{i}^{top2}$;
\textbf{2}. Split audios according to $sim_{i}^{top1}$ and $sim_{i}^{top2}$: $sim_{i}^{top1}>0.5$ and $sim_{i}^{top2}<0.4$ high-confidence; 
$sim_{i}^{top1}>0.5$ and $sim_{i}^{top2}>0.4$ median-confidence; 
$sim_{i}^{top1}<0.5$ low-confidence;
\textbf{3}. Use audio with median-confidence to find labels of samples, which comes from the same speaker but are given labels of multiple speakers by the GMVPG clustering algorithm, 
\textbf{4}. Two classes are merged into one when multiple models all show that they need to be merged.
\textbf{5}. Filter out audio that is low confidence, other audio is labeled by using predicted posterior probability.

\begin{algorithm}
    \SetAlgoNoLine 
    \SetKwFunction{FMain}{CheckCombine}
    \SetKwProg{Fn}{Function}{:}{}
    \Fn{\FMain{$U$}}{
        Calculate similarity between all $d$, for $d \in U$\;
        
        Using two-Gaussian distribution to fit the scores, $\mu_1$, $\sigma_1$, $w_1$, $\mu_2$, $\sigma_2$, $w_2$ represent the parameters of max and min Gaussian\;
        
        \eIf{$\mu_2>th_{nm}$ OR $w_1 >= 0.5$ OR $(\mu_1-\sigma_1) <= (\mu_2+\sigma_2) + \epsilon$}
        {\textbf{return} Yes}
        {\textbf{return} No}
    }
    \textbf{End Function}

    \SetKwFunction{FMain}{subspkCombine}
    \SetKwProg{Fn}{Function}{:}{}
    \Fn{\FMain{$E, G$}}{
        \For {each $e_{ij} \in E$}{
                $tmp_D = G_i2utts + G_j2utts$\;
                \If{$CheckCombine(tmp_D) == No$}
                {delete $e_{ij}$ from $E$}
            }  
        SCSG based on E to get G\;
    \textbf{return} G
    }
    \textbf{End Function}

    \SetKwFunction{FMain}{CbNewOld}
    \SetKwProg{Fn}{Function}{:}{}
    \Fn{\FMain{$G_{new}, G_{old}, E_{new.old}$}}{
            $U^{old} = G_{old}.nodes$;\\
            $U^{new} = G_{new}.nodes$;\\
            \For {each $u_{i} \in U_{new}$}{
                    $utt2subgraph_{i} = \{u^{old}_j\ if\ E_{new.old}(u_i,u^{old}_j)\}$ for $u^{old}_j \in U^{old}$ \;
                    $U_{i} = utt2subgraph_{i}.nodes$\;
                    \If {CheckCombine($U_{i}$) == No}
                    {delete $u_{i}$ from $G_{new}$ and $E_{new.old}$}
                }  
            SCSG based on $\{G_{new},E_{new.old},G_{old}\}$ to get G\;
        \textbf{return} G
    }
    \textbf{End Function}

\caption{Functions of GMVPG clustering\label{al4}}
\end{algorithm}

\subsection{Score calibration}
Score calibration has been an essential part in recent VoxSRCs. However, unlike track1\&2, track3 provided only 50 labeled speakers. To build a matched developing set as far as we can, we filtered 70 speakers with 20 segments each from unlabeled data according to their clustering purity. We gave labeled speakers more weight when generating developing trials as pseudo-label could be wrong. Finally, we got 40000 trials which is equal to the validation trials. The results show that our developing set is slightly easier but still brings performance gain. 

\subsection{Results}
Models shown in Table~\ref{t-base} are used to cluster. Systems in Table~\ref{t-s} are developed for Track3. Results we submitted are shown in Table~\ref{t-submit}.
After GMVPG clustering algorithm, we obtain 1711 speakers and 348861 utts.
After model training and label correction, we obtain 1760 speakers and 387700 utts.
\begin{table}[htbp]
  \centering
  \footnotesize
  \caption{Results of base model before/after adaptation}
    \begin{tabular}{ccccc}
    \toprule
    \multirow{2}[2]{*}{mdl} & \multirow{2}[2]{*}{loss} & \multirow{2}[2]{*}{vox2-train} & \multicolumn{2}{c}{t3-dev-EER} \\
          &       &       & ini   & adapt \\
    \midrule
    se-resnet34-32 & circle & clean-fb64-sgd & 16.86 & 14.29 \\
    cotnet & circle & clean-fb64-sgd & 16.65 &  14.55\\
    \midrule
    conformer & circle & aug-fb80-adam & 16.95 & 14.14 \\
    ecapa-large & circle & aug-fb80-adam & 18.02 & 14.62 \\
    se-resnet101-32 & circle & aug-fb80-adam & 14.06 & 11.90 \\
    \bottomrule
    \end{tabular}%
  \label{t-base}%
\end{table}%

\begin{table}[htbp]
  \centering
  \caption{Results of systems for Track3. v0 means pseudo labels before correction, and v1 means after.}
    \begin{tabular}{cccccc}
    \toprule
    \multirow{2}[2]{*}{} & \multirow{2}[2]{*}{mdl} & \multirow{2}[2]{*}{loss} & \multirow{2}[2]{*}{train} & \multicolumn{2}{c}{t3-dev-EER} \\
          &       &       &       & ini   & calib \\
    \midrule
    S1    & Res2Net50 & circle & v0-sgd & \textbf{8.45}  & 8.20 \\
    S2    & ResNet34 & circle & v0-sgd & 8.61  & \textbf{7.66} \\
    S3    & ECAPA-X4 & circle & v0-sgd & 9.57  & 8.81 \\
    S4    & ECAPA-X4 & circle & v0-adam & 10.47 & 9.65 \\
    S5    & ECAPA-X4 & circle & v1-sgd & 8.78  & 8.42 \\
    \bottomrule
    \end{tabular}%
  \label{t-s}%
\end{table}%

\begin{table}[htbp]
  \centering
  \caption{Results of systems that we submitted}
    \begin{tabular}{cccc}
    \toprule
    mdl   & mode  & dev-EER & eval-EER \\
    \midrule
    S1    & ini   & 8.45  & 8.07 \\
    S2    & ini   & 8.61  & 8.64 \\
    S1+S2 & ini   & 8.01  & 7.57 \\
    S1+S2+S3+S4 & ini   & 7.87  & 7.40 \\
    S1+S2+S3+S4+S5 & calib & \textbf{6.77}  & \textbf{7.03} \\
    \bottomrule
    \end{tabular}%
  \label{t-submit}%
\end{table}%

\section{Conclusions}

In this paper, we summarized our systems for VoxSRC2022 in detail. For track1, we explore various strong speaker embedding extractors and some training tricks. For Track3, we explore some domain adaptation methods firstly. Then, we propose a novel progressive sub-graph clustering algorithm based on two Gaussian fitting and multi-model voting to obtain pseudo labels. Thirdly, we explore some fine-tuning tricks to achieve better performance. Finally, we propose one label correction algorithm to correct noisy labels. Our Fusion systems achieve $6^{th}$ and $1^{st}$ place in Track 1 and 3 respectively.

\bibliographystyle{IEEEtran}

\bibliography{final}

\end{document}